\documentclass[12pt]{iopart}
\usepackage{iopams}
\newtheorem{lemma}{Lemma}

\newtheorem{theorem}{Theorem}
\newtheorem{corollary}{Corollary}

\def\tr{\mathop{\rm Tr}\nolimits}
\def\div{\mathop{\rm div}\nolimits}

\def\curl{\mathop{\rm curl}\nolimits}

\begin{document}

\title[gravito-magnetic spacetimes]
{Gravito-magnetic vacuum spacetimes: kinematic restrictions}

\author{Joan Josep Ferrando$^1$\ 
and Juan Antonio S\'aez$^2$}

\address{$^1$\ Departament d'Astronomia i Astrof\'{\i}sica, Universitat
de Val\`encia, E-46100 Burjassot, Val\`encia, Spain.}

\address{$^2$\ Departament de Matem\`atica Econ\`omico-Empressarial, Universitat de
Val\`encia, E-46071 Val\`encia, Spain}

\ead{joan.ferrando@uv.es; juan.a.saez@uv.es}

\begin{abstract}
We show that there are no vacuum solutions with a purely magnetic Weyl tensor with respect to an observer submitted to kinematic restrictions involving first order differential scalars. This result generalizes previous ones for the vorticity-free and shear-free cases. We use a covariant approach which makes evident that only the Bianchi identities are used and, consequently, the results are also valid for non vacuum solutions with vanishing Cotton tensor.
\end{abstract}

\pacs{0420C, 0420-q}

\submitto{\CQG}


\section{Introduction}

In an oriented spacetime $(V_4,g)$ of signature $\{ -, +,+,+ \}$, we can associate with any observer $u$ the electric and magnetic Weyl fields:
\begin{equation}
E= E[v] \equiv W(v;v) \, , \qquad \qquad H = H[v] \equiv *W(v;v)        \label{eh}
\end{equation}
where we denote $W(v;v)_{\alpha \gamma} = W_{\alpha \beta \gamma \delta} v^{\beta} v^{\delta}$. The electric and magnetic fields (\ref{eh}) determine the Weyl tensor fully and they play in general Relativity a similar role than the electric and magnetic fields play in electrodynamics \cite{bel}. Purely electric ($H=0$) gravitational fields are somewhat the relativistic version of Newtonian fields and wide and physically interesting classes of solutions are known. We can quote, for example, the static solutions and the other spacetimes admitting a normal shear-free observer \cite{tru}, as well as the silent universes \cite{marc}. Nevertheless, one knows few solutions with a purely magnetic Weyl tensor ($E=0$) (see references in  \cite{fsEM} \cite{van}) and some results are known that restrict their existence. Indeed, the magnetic irrotational dust models are subject to severe integrability conditions and it has been conjectured that only the Friedmann-Lemaître-Robertson-Walker models satisfy them \cite{maar2}. On the other hand, McIntosh {\it et al.} have shown that there are no vacuum solutions with a purely magnetic type D Weyl tensor, and they have conjectured that a similar restriction could take place for a wide class of type I spacetimes \cite{mcar}.

We can quote two significant steps in supporting the McIntosh {\it et al.} conjecture. First Haddow \cite{had} showed it when the observer is shear-free and, recently, Van der Bergh \cite{van} has showed the conjecture provided that the observer defines a normal congruence. In his proof, Van der Bergh apparently uses both, the Bianchi identities and the Einstein field equations, as well as the expression of the electric and magnetic Weyl fields obtained from the Ricci identities. On the other hand, Haddow obtained his result by using the Brans theorem \cite{brans} for the vacuum case. Here we improve these results in different aspects. Firstly we show that the conjecture is also true under weaker kinematic conditions on the observer. In fact these conditions involve first order differential scalars depending on the kinematic coefficients and they trivially hold when the shear or the vorticity vanish. Secondly, we use a covariant approach which makes evident that only the Bianchi identities are necessary and, consequently, these results are also valid for non vacuum solutions with vanishing Cotton tensor. Finally, we put forward that even more general results are known in type D metrics \cite{fsD} and that a similar generalization could also be obtained for type I spacetimes \cite{fsHom}.

\section{Gravito-magnetic observers in vacuum}

When the spacetime is purely magnetic with respect to an observer $u$, this observer defines a time-like Weyl principal direction. Then the magnetic Weyl tensor $H$ vanishes if, and only if, $\tr H^2=(H,H)=0$. In the 1+3 formalism relative to the observer $u$, when $E=0$ the Bianchi identities for the vacuum case (vanishing Cotton tensor) lead to the following restrictions on $H$ and on the shear $\sigma$, the vorticity vector $\omega$ and the acceleration $a$ of the observer $u$ \cite{maar}:
\begin{eqnarray}
[\sigma, H] = 3H(\omega)  \label{bi1} \\
\div H = 0   \label{bi2} \\
\curl H = - 2 a \wedge H  \label{bi3} 
\end{eqnarray}
where $\div$ and $\curl$ are, respectively, the covariant spatial divergence and curl operators, and $\wedge$ and $[\ , \ ]$ are the generalized covariant vector products (see for example \cite{maar} for more details). On the other hand, the Ricci identities give us the generic 1+3 expression for the magnetic field:
\begin{equation}  \label{Ricci-h}
H = \curl \sigma + \hat{D \omega} + 2 a \widehat{ \otimes} \omega
\end{equation}
where $D$ is the covariant spatial derivative and, for a tensor $A$, $\hat{A}$ means the projected trace-free symmetric part \cite{maar}.

From the Bianchi identity (\ref{bi1}) we have that the shear-free condition, $\sigma=0$, implies $H(\omega)=0$, so that, either the vorticity $\omega$ is zero or it is an eigenvector of $H$ with zero eigenvalue. The first condition leads to a normal shear-free congruence which is not compatible with $E=0$ as a consequence of the Trümper result \cite{tru}. The second condition implies that the Weyl tensor has a null eigenvalue in disagreement with the generalized Brans theorem \cite{brans} \cite{fsI}. 

Thus, we recover the Haddow result \cite{had}: {\it there are no vacuum solutions which are purely magnetic for a shear-free observer}. A similar result for a normal observer has been recently shown by Van der Bergh \cite{van}. But we have established our statement using exclusively the Bianchi identities and, consequently, it also applies for non vacuum solutions with a vanishing Cotton tensor. We will show below that the result by Van der Bergh can be extended in a similar way to Cotton-zero spacetimes. We will find this result as a corollary of our main statement that looks for stronger restrictions on the kinematic coefficients which are necessary in purely magnetic solutions.  

In order to demonstrate our theorem, we need two general properties of the 1+3 formalism that generalize, respectively, the divergence of a vector product and the cyclic quality of the scalar triple product. Namely, if $A,B$ are spatial trace-free symmetric tensors and $v$ a vector, one has the following relations:
\begin{eqnarray}
\div[A,B] = (\curl A, B)- (A, \curl B)     \label{pro1}\\
(A, v \wedge B) = - (v, [A,B])            \label{pro2}
\end{eqnarray}
We can calculate the divergence of $[\sigma, H]$ in two different ways. Making use of (\ref{bi1}) and taking into account (\ref{bi2}), one has:
\begin{equation}
\div[\sigma,H] =3 \div(H(\omega)) = 3(\div H, \omega) + 3 (D\omega,H) = 3(D\omega,H)  \label{c1}
\end{equation}
On the other hand, applying (\ref{pro1}) and using successively (\ref{bi3}), (\ref{Ricci-h}), (\ref{pro2}) and (\ref{bi1}), one obtains:
\begin{eqnarray}
\hspace{-1.5cm} \div[\sigma,H] & = & (\curl \sigma, H) - (\sigma, \curl H) = (H - D\omega - 2 a \otimes \omega, H) + 2 (\sigma, a \wedge H) =
\nonumber\\
& =  & \tr H^2 - (D\omega + 2 a \otimes \omega, H) - 2 (a , [\sigma, H]) =
\nonumber\\
& = & \tr H^2 - (D\omega,H) - 8 H(a, \omega)   \label{c2}
\end{eqnarray}
Then, taking into account expressions (\ref{c1}) and (\ref{c2}) we obtain the following restriction on the magnetic field $H$, the vorticity $\omega$ and the acceleration $a$:
\begin{equation}
\tr H^2 = 4 (H, D\omega + 2 a \otimes \omega)   \label{r1}
\end{equation}
On the other hand, the product by $H$ of equation (\ref{Ricci-h}) leads to a condition involving also the shear $\sigma$:
\begin{equation}
\tr H^2 = (H, \curl \sigma) + (H, D\omega + 2 a \otimes \omega)   \label{r2}
\end{equation}
The two conditions (\ref{r1}) and (\ref{r2}) allow us to state the following
\begin{lemma}  \label{lema1}
If a spacetime with vanishing Cotton tensor is purely magnetic with respect to an observer $u$, then the magnetic field $H$ and the kinematic coefficients associated with $u$ are submitted to the conditions:
\begin{eqnarray}
 \frac14 \tr H^2 =  (H, D\omega + 2 a \otimes \omega) = \frac13 (H, \curl \sigma)  \label{lema1a}
\end{eqnarray}
\end{lemma} 
This lemma leads to the following theorem.
\begin{theorem} \label{theo}
If a spacetime with vanishing Cotton tensor is purely magnetic with respect to an observer, then this observer is submitted to the conditions
\begin{equation}
(H, \curl \sigma) > 0 \, ,  \qquad  \qquad  (H, D\omega + 2 a \otimes \omega) > 0    \label{theo1}
\end{equation}
In particular, there are no vacuum solutions for which the Weyl tensor is purely magnetic with respect to an observer satisfying $(H, \curl \sigma) \leq 0$ or $(H, D\omega + 2 a \otimes \omega) \leq 0$.
\end{theorem}
In the previous theorem we find first order differential conditions on the vorticity $\omega$ or on the shear $\sigma$  which are identically satisfied when $\omega=0$ or $\sigma =0$, respectively. Nevertheless, we also have algebraic conditions for a rotating or a shearing observer that are not compatible with the purely magnetic condition. Indeed, from equation (\ref{bi1}) and the generalized Brans theorem \cite{brans} \cite{fsI}, it follows:
\begin{corollary} \label{cor1}
There are no vacuum (Cotton-zero) solutions for which the Weyl tensor is purely magnetic with respect to an observer satisfying $[H, \sigma] = 0$ or $H(\omega) = 0$.
\end{corollary}

In theorem \ref{theo} and corollary \ref{cor1} we can find conditions that restrict the purely magnetic spacetimes. These conditions are linear on the shear $\sigma$ or on the vorticity $\omega$ and they also involve the magnetic Weyl field $H$. Nevertheless, we can remove $H$ in some of these conditions and we find quadratic ones on the kinematic coefficients. Indeed, if we substitute expression (\ref{Ricci-h}) for $H$ in (\ref{lema1a}) we obtain a different version of lemma \ref{lema1}:
\begin{lemma}
If a spacetime with vanishing Cotton tensor is purely magnetic with respect to an observer $u$, then the magnetic field $H$ and the kinematic coefficients associated with $u$ are submitted to the conditions:
\begin{eqnarray}
 \frac12 \tr H^2 = \tr (\curl \sigma)^2 - \tr (\hat{D \omega} + 2 a \widehat{ \otimes} \omega)^2  \, ,  \label{lema2a} \\
\tr(\curl \sigma)^2 - 3 \tr (\hat{D \omega} + 2 a \widehat{ \otimes} \omega)^2 = 2(\curl \sigma, \hat{D \omega} + 2 a \widehat{ \otimes} \omega)   \label{lema2b}
\end{eqnarray}
\end{lemma} 
This lemma allows us to state the restrictions in theorem \ref{theo} using exclusively the kinematic coefficients of the observer.
\begin{theorem} \label{theo2}
If a spacetime with vanishing Cotton tensor is purely magnetic with respect to an observer, then this observer is submitted to the condition (\ref{lema2b}) and the restrictions
\begin{equation}
0 < \tr (\hat{D \omega} + 2 a \widehat{ \otimes} \omega)^2 <  \tr (\curl \sigma)^2 
\end{equation}
In particular, there are no vacuum solutions for which the Weyl tensor is purely magnetic with respect to an observer satisfying one of the following conditions:
\description
\item (i) $\tr(\curl \sigma)^2 - 3 \tr (\hat{D \omega} + 2 a \widehat{ \otimes} \omega)^2 \not= 2(\curl \sigma, \hat{D \omega} + 2 a \widehat{ \otimes} \omega)$   
\item (ii) $0 \geq \tr (\hat{D \omega} + 2 a \widehat{ \otimes} \omega)^2$ 
\item (iii) $\tr (\hat{D \omega} + 2 a \widehat{ \otimes} \omega)^2 \geq \tr (\curl \sigma)^2$
\end{theorem}

\section{Concluding remarks}

The results in this work bring a new improvement in supporting the conception that purely magnetic vacuum solutions could not take place. In a such spacetime the observer, with a vanishing relative electric Weyl tensor and a non vanishing magnetic one, would be submitted to the kinematic restrictions given in theorem 1 and theorem 2. So, in looking for a counterexample for the McIntosh {\it et al.} conjecture we must impose that a principal observer satisfies these conditions. On the other hand, it is worth pointing out that the restrictions on the purely magnetic vacuum solutions also apply on the non vacuum ones with vanishing Cotton tensor.

We have done no hypothesis on the non degenerate character of the Weyl tensor and our results apply on both Petrov types that are compatibles with the magnetic condition, type D and type I metrics. But for the vacuum type D spacetimes there are no magnetic solutions independently of the kinematic properties of the observer \cite{mcar}. A proof of this result using only Bianchi identities has been obtained elsewhere \cite{fsD} and so it is also valid for non vacuum Cotton-zero solutions. On the other hand, in this last work we show that not only the purely magnetic solutions are forbidden, but also a wider class of type D solutions. More precisely, we have shown \cite{fsD}: {\it if a spacetime with vanishing Cotton tensor has a type D Weyl tensor with eigenvalues of constant argument, then it is a purely electric solution}. This means that the constant argument takes, necessarily, the values $0$ or $\pi$. 

A question arises in a natural way: is there a similar generalization for algebraically general spacetimes?, that is, is it possible to extend to a wider class of type I solutions the restrictions that we have obtained here for the purely magnetic ones? Is this class related with the type I metrics whose Debever null principal directions span a 3--plane as McIntosh {\it et al.} seem to suggest \cite{mcar}? We will analyze all these questions in detail elsewhere \cite{fsHom} but we here advance what kind of results we are obtaining. We show that the four Debever direction span a 3--plane if, and only if, the electric and magnetic Weyl fields relatives to the principal observer are proportional, $\alpha E + \beta H = 0$. This property always holds for type D metrics with respect to any principal observer. In this case the eigenvalue has constant argument only when $E$ and $H$ are homothetic. And we could generalize the results that we have shown in this article for type I spacetimes verifying this condition \cite{fsHom}: {\it when a metric with vanishing Cotton tensor has homothetic electric and magnetic Weyl fields with respect to an observer satisfying one of the conditions in theorems 1 and 2, then the spacetime is purely electric}.

\ack
This work has been partially supported by the Spanish Ministerio de Ciencia y Tecnolog\'{\i}a, project AYA2000-2045.

\end{document}